\documentclass[11pt]{article}
\usepackage{epsfig}
\usepackage{amssymb}
        \newdimen\eqskip
        \newdimen\txtskip
        \baselineskip=\txtskip
        \newdimen\mysep                
        \newdimen\hmysep
\begin{document}
       
  \newcommand{\ccaption}[2]{
    \begin{center}
    \parbox{0.85\textwidth}{
      \caption[#1]{\small{{#2}}}
      }
    \end{center}
    }
\newcommand{\BS}{\bigskip}
% MATH SYMBOLS
\def    \be             {\begin{equation}}
\def    \ee             {\end{equation}}
\def    \ba             {\begin{eqnarray}}
\def    \ea             {\end{eqnarray}}
\def    \nn             {\nonumber}
\def    \=              {\;=\;}
\def    \frac           #1#2{{#1 \over #2}}
\def    \ret            {\\[\eqskip]}
\def    \ie             {{\em i.e.\/} }
\def    \eg             {{\em e.g.\/} }
\def \lsim{\mathrel{\vcenter
     {\hbox{$<$}\nointerlineskip\hbox{$\sim$}}}}
\def \gtrsim{\mathrel{\vcenter
     {\hbox{$>$}\nointerlineskip\hbox{$\sim$}}}}
\def    \bentarrow      {\:\raisebox{1.1ex}{\rlap{$\vert$}}\!\rightarrow}
\def \sss {\scriptscriptstyle}
% UNITS                 
\def    \kev            {\mbox{$\mathrm{keV}$}}
\def    \mev            {\mbox{$\mathrm{MeV}$}}
\def    \gev            {\mbox{$\mathrm{GeV}$}}
\def    \ifb            {\mbox{$\mathrm{fb}^{-1}$}}

% OTHERS
\def\mh{$m_H$}
\def\azero{{\mbox{$a^0$}}}
\def\ma{{\mbox{$m_{a^0}$}}}
\def\masq{{\mbox{$m^2_{a^0}$}}}
\def\oneS{\Upsilon_{1S}}
\def\mups{M_\Upsilon}
\def\mupssq{M^2_\Upsilon}
\def\moneS{M_{1S}}
\def\gee{\Gamma_{e^+e^-}}
\def\bee{B_{e^+e^-}}
\def\gtot{\Gamma_\mathrm{tot}}
\def \avsig{\langle \sigma \rangle}
\def\thetag{\theta_\gamma}
\def\costhg{\cos\theta_\gamma}
\def\mumu{\mu^+\mu^-}
\def\epem{e^+e^-}
\def\mmumu{M_{\mu\mu}}
\def\mmu{m_{\mu}}
\def\mmusq{m^2_{\mu}}
%%%%%%%%%%%%%%%%%%%%%%%%%%%%%%%%%%%%%%%%%%%%%%%%%%%%%%%%%%%%%%%%%%%%%%
%\begin{titlepage}
%\nopagebreak
{\flushright{
        \begin{minipage}{5cm}
	Bicocca-FT-07-6 \\
        CERN-PH-TH/2007-062\\
        {\tt hep-ph/yymmnnn}
        \end{minipage}        }
        
}
%\vfill
\vskip 1.2cm
\begin{center}
{ { \bf \sc \Large
Radiative quarkonium decays and
the NMSSM Higgs  \\[0.5cm] interpretation
of the HyperCP $\Sigma^+ \to
p\mu^+\mu^-$ events}}
\\[1cm]
%\vfill                                                       
{\bf Michelangelo L. MANGANO$^{(a)}$ and Paolo NASON$^{(b)}$}
\\[0.5cm]
{\small
$^{(a)}$ CERN, PH-TH, Geneva, Switzerland \\
$^{(b)}$ INFN, Sezione di Milano Bicocca, Italy
}
\end{center}                                   
%\nopagebreak
%\vfill
%\vskip 3cm
\begin{center}
{\bf Abstract}
\end{center}
We study the potential of radiative decays of the $\oneS$ and of the
$\phi$ mesons to search for a light pseudoscalar Higgs boson,
proposed as a possible interpretation of  $\Sigma^+ \to
p\mu^+\mu^-$ events observed by the HyperCP collaboration at
Fermilab. We conclude that the detection of this signal should
certainly be
possible with the current CLEO $\oneS$ data, and is within the reach
of KLOE in at least part of the range of couplings suggested by
the HyperCP findings.

%\vskip 1cm
%CERN-TH/2006-XXX\hfill \\
%\today \hfill  
%\vfill       
%\end{titlepage}

\section{Introduction}
\label{sec:intro}
The HyperCP collaboration at Fermilab has recently
reported~\cite{Park:2005ek} the observation of 3 $\Sigma^+\to p \mumu$
decays, with the dimuon invariant mass $\mmumu=214.3\pm0.5$~MeV,
consistent with the decay of a narrow neutral intermediate state
$P^0$. Among the several candidate new particles put forward to
explain this observation~\cite{He:2005we}-\cite{He:2006fr}, 
the possibility~\cite{He:2006fr} that $P^0$
is a light pseudoscalar Higgs ($\azero$) 
in the next-to-minimal supersymmetric Standard Model
(NMSSM) interestingly fits within theoretical frameworks recently
explored in the attempt to loosen the very tight constraints set in
the MSSM by the non-observation of the Higgs at LEP2. 

Comprehensive studies of the implications of a light NMSSM Higgs for
current and future measurements have been performed. Excellent
examples are given, for example, in
\cite{Dermisek:2006py,Hiller:2004ii} and in a
comprehensive review of non-standard Higgs scenarios~\cite{Accomando:2006ga}. 
Here we limit ourselves to explore one important
direct implication of the proposal made in~\cite{He:2006fr}. Namely
the possibility that $\azero$ be produced in quarkonium radiative
decays $V\to\gamma\azero$~\cite{Wilczek:1977zn}, 
where, in our analysis, $V=\oneS$ or
$\phi(1020)$.  The most recent and precise limit on $\oneS\to
\gamma\azero$ was given by CLEO~\cite{Balest:1994ch}, where however
$\azero$ is required to be stable and non-interacting. While several
studies with visible Higgs final states have been performed since the
early works of~\cite{Franzini:1987pv}, we are not
aware of limits valid in the case of prompt $\azero\to\mumu$ decays
in the range of branching ratios B($\oneS\to\gamma\azero$) relevant to
the model of~\cite{He:2006fr}. Likewise, we are not aware of
experimental studies on the potential of $\phi$ decays to detect or
constrain such scenarios. The excellent performance of the KLOE
detector at DA$\Phi$NE (see e.g.~\cite{Ambrosino:2007nx}), 
and the prospects for a continued operation at much
higher luminosities\footnote{See {\tt
    http://www.lnf.infn.it/lnfadmin/direzione/roadmap/roadmap.html }}, 
make the consideration of $\phi$ decays extremely
interesting.

 Radiative quarkonium decays have been studied already in
the context of other hypothetical interpretations of the HyperCP
findings, namely the identification of $P^0$ with a
sgoldstino~\cite{Demidov:2006pt}. In that case decay rates
are typically beyond the present experimental sensitivities; in the
case of a pseudoscalar Higgs, as we shall show, decay rates consistent
with the HyperCP data are well within the reach of current experiments.

The virtue of our analysis is that it is only based on the assumptions
on the coupling of $\azero$ to down-type quarks and to leptons, and it
is independent of the details of the pattern of Higgs expectation
values and mixings which need to be fixed to study other possible
implications of this class of
models~\cite{Dermisek:2006py,Hiller:2004ii}. 
Furthermore, the quarkonium decay does
not depend on the size of FCNC couplings of $\azero$, which enter in
the analysis of the $\Sigma^+\to p\mumu$ decay. Finally, the
prediction for $\phi$ decays only depends on the $\azero$ coupling to
the $s$ quark, which is what is probed in $\Sigma^+\to p\mumu$, and
therefore evades the potential loophole of the $\oneS$ study, namely
that in more complex Higgs sectors the coupling to the $b$ quark could
be independent of the coupling to the $s$. Therefore, independently of
the confirmation or invalidation of the hypothesis
of~\cite{He:2006fr}, our note should be taken as an encouragement to
revive the exploration of radiative quarkonium decays as a powerful
tool to detect or constrain non-standard Higgs models, extending the
considerations presented in ref.~\cite{Dermisek:2006py}. Other
motivations to continue the experimental study of quarkonium radiative
decays as a probe for new physics are presented
in~\cite{Sanchis-Lozano:2006gx}. 

We start by briefly summarizing the key couplings of $\azero$ to the
fermion sector. 
\be
L_{a^0f\bar{f}} = -(g_u \, m_u \, \bar{u}\gamma_5 u +
g_d \, m_d \, \bar{d}\gamma_5 d + g_d \, m_\ell \,\bar{\ell} \gamma_5
\ell) 
\; \frac{i \, a^0}{v} \; ,
\ee
where $v\sim 246$~GeV, and the couplings $g_{u,d}$ depend on the
details of the SUSY and EW symmetry breaking pattern. 
The analysis the HyperCP data in~\cite{He:2006fr}
suggests $g_d={\cal O}(1)$, compatible with the constraint $\vert g_d
\vert \lsim 1.2$ from the muon $g-2$~\cite{He:2005we}. A rough lower limit
on $g_d$ can be obtained by requiring the Higgs lifetime
to be short enough to guarantee the reconstruction of a common vertex
for the three tracks relative to the muons and the proton in the
HyperCP data. Using
\be
\Gamma(a^0 \to \mu^+\mu^-) = \frac{g_d^2}{8\pi}\frac{\mmusq}{v^2}\;
\left[ m_{a^0}^2-4\mmusq \right]^{1/2} \;
\ee
and using a Lorentz $\gamma$ factor of about 230 (the typical momentum
of the dimuon pair in the HyperCP data is 50~GeV~\cite{hpark}), the
average decay length  of the $a^0$ in the laboratory frame turns out
to be $\sim 0.02$~cm/$g_d^{2}$. Requiring this not to exceed a
 resolution in $z$ of the order of 60~cm~\cite{hpark}, 
would lead to $g_d \gtrsim 0.02$.

The above couplings induce quarkonium decays with a rate given by:
\be
\frac{B(V\to\gamma\azero)}{B(V\to\epem)} = 
\frac{G_F \, m_Q^2}{\sqrt{2}\pi\alpha} \; g_d^2 \;
(1-\frac{\masq}{m_V^2}) \, F \sim 3.6\times10^{-4} 
\frac{m_Q^2}{\mathrm{GeV}^2} F g_d^2
\ee
where $m_Q$ is the heavy quark mass and 
$F\sim 0.5$ includes the effect of QCD radiative corrections.
For our numerical analysis, in the case of the $\oneS$ decays we'll use 
$m_b=\moneS/2$. In the case of the $\phi$, the small mass leads to a
rather large uncertainty on the
choice of the quark mass and of the size of QCD
radiative corrections. As in the case of the $\oneS$, we shall adopt the
convention $m_s=m_\phi/2$, and discuss in appendix \ref{app:QCD}
the motivations and limitations of this choice.
With these assumptions, and using $B(\oneS\to\epem)=2.4\times 10^{-2}$ and
$B(\phi\to\epem)=3\times 10^{-4}$, we obtain:
\be
B(\oneS\to \gamma\azero) = 1.9 \times 10^{-4} F g_d^2 \quad , \quad
B(\phi \to \gamma\azero) = 2.8 \times 10^{-8} F g_d^2
\ee
The rate of $\oneS$ produced at the B factories depends on the
details of the beam energy resolution, which is typically larger than
the intrinsic $\oneS$ width. Using the CLEO data discussed in
~\cite{Besson:2005ud} as a benchmark, we
extract a production rate of $21\times 10^6/1.13\ifb \sim 1.8\times
10^{7}$fb. A similar rate is expected for BaBar and Belle, due to
comparable beam energy profiles. The DA$\Phi$NE beams allow instead to
operate at the peak of the $\phi$ resonance, with a rate given by:
\be 
\sigma(e^+e^- \to \phi) = \frac{12\pi}{m_\phi^2} \; \bee \; A_{\mathrm ISR}
\;\sim \; 2.8 \times 10^9\, \mathrm{fb}  \; ,
\ee
where $A_{\mathrm ISR}\sim 2/3$ is the QED ISR correction factor.
This leads to the following estimates for the number of
$V\to\gamma\azero\to\gamma\mumu$ decays per \ifb (assuming $F=1/2$):
\ba
N(\oneS\to \gamma\azero \to\gamma\mumu) &=& 1700 \; g_d^2 \\
N(\phi \to \gamma\azero \to\gamma\mumu) &=& 40 \; g_d^2
\; .
\ea
Considering that the current KLOE dataset consists of about 2.5\ifb, and
that a factor of $>10$ increase is planned for the future, we conclude
that both the B factories and DA$\Phi$NE are potentially sensitive to the
range of couplings required for the Higgs interpretation of the
HyperCP events. 

{\renewcommand{\arraystretch}{1.2}
\begin{table}
\begin{center}
\begin{tabular}{|l|c|} 
\hline
$\phi$,  213.8 MeV$<\mmumu<$ 214.8 MeV &   0.5~pb   \\ 
$S/\sqrt{B}$ for 2.5 \ifb                &   1.8 $g_d^2$  \\
\hline
$\phi$,  214.2 MeV$<\mmumu<$ 214.4 MeV &   0.1~pb   \\ 
$S/\sqrt{B}$ for 2.5 \ifb                &   3.9 $g_d^2$  \\
\hline
$\oneS$, $2m_\mu <\mmumu<$ 220 MeV     &  0.05~pb  \\ 
$S/\sqrt{B}$ for 1 \ifb                & 150 $g_d^2$   \\
\hline
\end{tabular}
\ccaption{}{\label{tab:xsec} Cross section for the QED process
  $e^+e^- \to \mu^+\mu^- \gamma$, and significance of the $\azero$ signal.}
\end{center}                                         
\end{table} }

The actual observability of this signal depends however on the
background rates. While the signature is rather sharp (a monochromatic
photon, produced at large angle with energy close to $m_V/2$ and
recoiling against a low-mass dimuon pair), a potentially large
background can arise from the continuum QED production of $\epem \to
\gamma\mumu$ events. The actual size of the background will depend on
the dimuon mass resolution, where the signal has a sharp peak with an
intrinsic width below the MeV.  We point out that, for a muon pair
produced near threshold, like in the case of the HyperCP signal, the
invariant mass measurement should have an improved sensitivity. The
invariant mass (as a function of the angular separation and the
momentum of the muons) is in fact near a minimum, and therefore has a
reduced sensitivity to the measured parameters and their errors, as
discussed in an Appendix \ref{app:kinthr}. For our numerical study we
assume dimuon mass resolutions equal to $\pm 5$~MeV for the $\oneS$, and $\pm
0.5$~MeV or $\pm 0.1$~MeV for the $\phi$.

 The resulting background rates are given in table~\ref{tab:xsec},
together with an estimate of the statistical significance of a
possible signal.  Here we required the photon to have energy larger
than 1 GeV (100 MeV) for the $\oneS$ ($\phi$), and to
 be emitted at an
angle with respect to the beam such that $\vert \cos\theta \vert <
0.7$.  Since the photon distribution in the signal is proportional to
$1+\cos^2\theta$, this cut has a signal efficiency of about 0.61\footnote{We
verified that $S/\sqrt{B}$ is rather constant, and maximal, when the cut on
$\cos\theta$ is varied in the region 0.6--0.8.}.

The separation of the signal from the background appears rather
clear at the $\oneS$ with 1\ifb, even if the coupling is significantly
smaller than its preferred ${\cal O}(1)$ value. 
The case for KLOE with its current
statistics is at the borderline, even though there is a margin of
uncertainty due to the large size of the QCD corrections. A mass
resolution of 0.1~MeV appears required to yield a significant signal.

Higher statistics, particularly at KLOE, would
allow to strenghten the constraints on similar explorations of exotic
BSM scenarios.

{\bf Acknowledgments} We thank D. Asner, S. Bertolucci, F. Bossi,
P. Dauncey, G. Isidori and H. Park for useful discussions and
contributions to this study. 

\appendix 
\section{Background estimates}
The exact matrix element for the process $e^-(p_1) e^+(p_2) \to \gamma(k)
\mu^-(q_1)\mu^+(q_2)$ can be easily calculated. We show here the
result in the case of initial-state emission only, which is the
relevant one for our signal:
\def\pk{(p_1 k)}
\def\pbk{(p_2 k)}
\def\pq{(p_1 q_1)}
\def\pqb{(p_1 q_2)}
\def\pbq{(p_2 q_1)}
\def\pbqb{(p_2 q_2)}
\def\pQ{(p_1 Q)}
\def\pbQ{(p_2 Q)}
\ba
\overline{\sum}_{pol} \vert M(\epem \to \gamma \mumu) \vert ^2 &=&
  4\frac{e^6}{\pk\pbk} \; \frac{1}{Q^2} \; \left[ \pq^2+\pbq^2
  +\pqb^2+\pbqb^2 + \right . \nn \\
&& \frac{2\mmusq}{Q^2} \left . \left(\pQ^2 + \pbQ^2 \right) \right]
\ea
where  $q_1+q_2=Q$.
The final-state kinematics of the signal, two low-mass muons recoiling
against the photon, allows us to approximate the cross section as:
\be
\frac{d\sigma(\gamma\mumu)}{d\cos\theta_\gamma} = 
\frac{d\sigma(\gamma\gamma)}{d\cos\theta_\gamma} \times
\frac{dQ^2}{Q^2} \frac{\alpha}{2\pi} P(z) \, dz  \; ,
\ee
where
\be \label{eq:gammagamma}
\frac{d\sigma(\gamma\gamma)}{d\cos\theta_\gamma} =
2e^4\frac{u^2+t^2}{ut} \; \frac{1}{32 \pi s} \; ,
\ee
and
\be
P(z) = [z^2+(1-z)^2+2\frac{\mmusq}{Q^2}] \; 
\ee
is the Altarelli-Parisi splitting function, for
\be q_1^\mu = z Q^\mu, \quad \frac{1}{2}(1-\sqrt{1-\frac{4\mmusq}{Q^2}}) <
z < \frac{1}{2}(1+\sqrt{1-\frac{4\mmusq}{Q^2}}) \; .
\ee
After integrating over $z$, over the dimuon mass in the range
$2\mmu+\Delta_1<Q<2\mmu+\Delta_2$ (with $\Delta_{1,2}<<\mmu$), and over the
photon angular distribution in the
range $-c<\cos\theta<c$, we obtain:
\be \label{eq:approx}
\sigma(\gamma\mumu) = \frac{4\pi\alpha^2}{s}\left(
\log\frac{1+c}{1-c}-c\right)
\frac{\alpha}{3\pi} \frac{\Delta_2^{3/2}-\Delta_1^{3/2}}{\mmu^{3/2}} \; .
\ee
While the results of table~1 were obtained using the exact formula, 
eq.~(\ref{eq:approx})
 is an excellent approximation to the exact result, which can be
used to easily explore the background size under cuts different from
those used in table~1.

\section{QCD corrections}\label{app:QCD}
The QCD corrections to the $V\to P+\gamma$
decay of quarkonium are given by \cite{Nason:1986tr}
\begin{equation}
  \label{eq:main} \frac{\Gamma (V \rightarrow P + \gamma)}{\Gamma (V
  \rightarrow e^+ e^-)} = \frac{G_F  m_q^2}{\sqrt{2} \pi
    \alpha} (1 - m_P^2 / m_V^2) \left(1 - \alpha_s C_f / \pi \left[ a_P \left(\frac{E_\gamma}{E^{\rm max}_\gamma}\right) - \frac{1}{4} \right]\right)
\end{equation}
where we have assumed for simplicity $g_d=1$.
The $-1/4$ term accounts for the QCD corrections to the $e^+ e^-$ decay rate.
This result is obtained by renormalizing the quark mass in the on-shell scheme.
Thus, the quark mass appearing in Wilczek formula is the pole mass; it
can be given in terms of the $\overline{\rm{MS}}$ mass evaluated at the scale
of the quark mass itself as \cite{Sachrajda2006}
\begin{equation}
  m_q = \bar{m}_q ( \bar{m}_q) \left[ 1 + C_f \alpha_s / \pi \right]
\end{equation}
so that eq. (\ref{eq:main}) becomes
\begin{equation}
  \label{eq:vhgmsbmass} \frac{\Gamma (V \rightarrow P + \gamma)}{\Gamma (V
  \rightarrow e^+ e^-)} = \frac{G_F \bar{m}_q^2 ( \bar{m}_q)}{\sqrt{2}
    \pi \alpha} (1 - m_P^2 / m_V^2) \left(1 - \alpha_s C_f / \pi
    \left[ a_P \left(\frac{E_\gamma}{E^{\rm max}_\gamma}\right) - 2 -
    \frac{1}{4} \right]\right),
\end{equation}
where $a_P$ is an increasing function of its argument, with a maximum
value $a_P (1) \approx 6.62$. The strange $\overline{\rm{MS}}$ mass is
determined to be around 0.1 GeV at 2 GeV \cite{Sachrajda2006}.  Its
value at a scale equal to the mass itself can be obtained using the
renormalization group equation relation
\begin{equation}
  \label{eq:runmass} \frac{\bar{m}_q (\mu^2)}{\bar{m}_q (\mu_0^2)} = \left(
  \frac{\alpha_s (\mu^2)}{\alpha_s (\mu_0^2)} \right)^{4 / 9},
\end{equation}
and since $\alpha_s (\mu^2)$ diverges for $\mu < 0.5$ GeV, one would
always determine $\bar{m}_s ( \bar{m}_s) > 0.5$ GeV. Using 4-loop
running for the strange mass \cite{Vermaseren:1997fq}, $\alpha_s (M_Z)
= 0.119$ and $\bar{m}_s (2 {\rm GeV}) = 0.1 {\rm GeV}$, we find
$\bar{m}_s ( \bar{m}_s) = 0.6 {\rm GeV}$.  The next question is what
value of $\alpha_s$ one should use in eq. (\ref{eq:vhgmsbmass}). The
coefficient of $\alpha_s$ is $\approx 1.85$, and the scale of
$\alpha_s$ should be taken around the annihilation energy.  We find
$\alpha_s(1\,{\rm GeV}) \approx 0.5$, so that at this scale the
radiative corrections has the same size of the Born term. Assuming
that the correction exponentiates, we get $\exp (- 0.5 \times 1.85)
\approx 0.4$.  Assuming instead a rational form $1 / (1 + 0.5 \times
1.85) \approx 0.5$.  We thus conclude that setting the strange mass to
half of the $\phi$ mass, and using a correction factor of order of a
half yields a reasonable estimate. We should not forget, however, that
uncertainties are very large, so that a rate a few times larger or
smaller than our estimate is not unconceivable.

\section{$\mu^+\mu^-$ kinematics near threshold}\label{app:kinthr}
The invariant mass of the two muons, as a function of their momenta
$p_1$, $p_2$ and thair angular separation $\theta$ is
\begin{eqnarray}
M_{\mu\bar\mu}^2&=&4m_\mu^2+2p_1 p_2 (1-\cos\theta)+2 [E_1 E_2
-p_1 p_2 -m_\mu^2] \nonumber \\\label{eq:pairmass} &=&4m_\mu^2+2p_1
p_2 (1-\cos\theta)+2m_\mu^2 \frac{(p_1-p_2)^2}{E_1 E_2 + p_1 p_2
+m_\mu^2}\,.
\end{eqnarray}
Near threshold, the second and third terms in eq.~(\ref{eq:pairmass})
become small, so that in this limit
\begin{equation}\label{eq:pairmassappr}
M_{\mu\bar\mu}\approx 
2m_\mu+\frac{p^2}{4m_\mu} \theta^2+\frac{m_\mu}{4}\frac{\Delta p^2}{E^2}\,,
\end{equation}
where $p$ is the common value of $p_1$ and $p_2$, $E$ is their energy,
and $\Delta p=p_1-p_2$ is their small momentum difference. Experimental
errors $\delta\theta$ and $\delta \Delta p$ on the measurement of the
angle and the momentum affect the invariant mass measurement by
\begin{equation}
\delta M_{\mu\bar\mu}\approx \frac{p^2}{2m_\mu} \theta\delta \theta
+\frac{m_\mu}{2}\frac{\Delta p}{E^2} \delta \Delta p\,.
\end{equation}
Thus the error is reduced by the small factors $\theta$ and $\Delta p$.

\end{document}

%\begin{figure}
%\begin{center}
%\includegraphics[width=0.45\textwidth,clip]{moneS.eps}
%\hfil
%\includegraphics[width=0.45\textwidth,clip]{mphi.eps}
%\ccaption{}{\label{fig:mass} Dimuon mass spectrum (pb/bin).}
%\end{center}
%\end{figure}
%\begin{figure}
%\begin{center}
%\includegraphics[width=0.45\textwidth,clip]{thetaoneS.eps}
%\hfil
%\includegraphics[width=0.45\textwidth,clip]{thetaphi.eps}
%\ccaption{}{\label{fig:theta} Angular separation between the muons (pb/bin).}
%\end{center}
%\end{figure}
%\clearpage